\documentclass[12pt]{article}
\usepackage{amssymb,amsmath}
\usepackage[noblocks]{authblk}
\usepackage[top=0.75in, bottom=0.75in, left=0.75in, right=0.75in, dvips]{geometry}
\usepackage{caption}
\begin{document}
\textwidth 10.0in 
\textheight 9.0in 
\topmargin -0.60in
\title{Renormalization Scheme Dependence in\\
$b \rightarrow u\ell^-\overline{\nu}_\ell$ Semileptonic Decays}
\author[1,2]{D.G.C. McKeon}
\affil[1] {Department of Applied Mathematics, The
University of Western Ontario, London, ON N6A 5B7, Canada} 
\affil[2] {Department of Mathematics and
Computer Science, Algoma University, \newline Sault St.Marie, ON P6A
2G4, Canada}
\date{}
\maketitle    

\maketitle
\noindent
email: dgmckeo2@uwo.ca

\begin{abstract}
After reviewing how the renormalization group equation can be used to sum logarithmic corrections to the decay rate $\Gamma$ for the process $b \rightarrow u\ell^-\overline{\nu}_\ell$ when using minimal subtraction, we consider renormalization scheme dependence for this calculation when employing this  renormalization scheme.  In this calculation, an ambiguity resides in the running strong coupling and in the running $b$ quark mass.  The ambiguity usually associated with the  renormalization mass scale $\mu$ is shown to cancel. It is demonstrated how in one renormalization scheme, there are only leading-log contributions to $\Gamma$.  Another choice of renormalization scheme results in $\Gamma$ being expressed in terms of the two-loop contribution to the $\beta$-function associated with the strong coupling and the one-loop contribution to the anomalous mass dimension as well as a set of renormalization scheme invariant parameters.
\end{abstract}

\section{Introduction}

Excising the divergences that arise in perturbative calculations in quantum field theory necessitates a renormalization procedure that leads to ambiguities when one works to finite order [1]. The mass scale $\mu$ that arises in the course of renormalization is one such ambiguity; this leads to the renormalization group (RG) equation, which permits one to sum all logarithmic corrections arising in perturbation theory.  There are additional ambiguities, even when using a mass independent renormalization scheme (RS) [2,3].  They can be parameterized by the expansion coefficients for the $\beta$-function associated with the coupling [4] beyond two-loop order and the expansion coefficients for the anomalous mass dimension  beyond one-loop order [5].

In ref. [6] the RG equation has been used to perform the leading-log (LL), next-to-leading-log (NLL) etc. summations of perturbative contributions to the process $R(e^+e^- \rightarrow$ hadrons).  There it is shown that by including up to the N$^{(3)}$LL order in $R$, dependence on $\mu$ is greatly reduced when using the $\overline{MS}$ RS [7] and the four-loop perturbative corrections to $R$ and to the $\beta$-function associated with the strong coupling.  This is not surprising, as the exact expression for $R$ is independent of $\mu$, and by performing the summation of logarithms to all orders, a better approximation to this exact result should arise.

In ref. [8], the RS dependence of $R$ is examined and it is shown how by choosing a particular scheme, it is possible to express $R$ in terms of RS invariants and a two-loop coupling; in another RS, $R$ is only quadratic in the coupling. In this paper, we will extend this discussion of RS dependence to a process in which there is an additional ambiguity due to mass dependence; in particular we will consider the amplitude $\Gamma$ for the semi-leptonic $b$ quark decay $b \rightarrow u\ell^-\overline{\nu}_\ell$.  We will show that even when using a mass independent RS, by combining RG summation with a judicious choice of RS, it is possible to express $\Gamma$ in terms of RS invariants and the pole mass of the $b$ quark.  Furthermore, dependence on the renormalization scale parameter $\mu$ is replaced by dependence on the value of the running coupling and the running mass at some mass scale.  One choice of RS has $\Gamma$ being reduced to a LL sum, another choice has the running coupling and mass expressed in closed form.

\section{Renormalization Scheme Dependency in $b$-Quark Decays}

A perturbative evaluation of the amplitude for the decay $b \rightarrow u\ell^-\overline{\nu}_\ell$ leads to the expression
\begin{equation}
\Gamma = [m(\mu)]^5 \sum_{n=0}^\infty \sum_{k=0}^n T_{nk} a^n(\mu)\ln^k\left(\frac{\mu}{m}\right)
\end{equation}
where $m(\mu)$ is the running mass for the $b$ quark and $a(\mu)$ is the running strong coupling. (We assume five active quark flavours and have absorbed an overall factor of $\frac{G_F^2|V_{ub}|^2}{192\pi^3}$ into the expansion coefficients $T_{nk}$.)  As $\Gamma$ is independent of the renormalization scale parameter $\mu$, we have the RG equation
\begin{equation}
\mu\frac{d\Gamma}{d\mu} = 0 = \left( \mu\frac{\partial}{\partial\mu} + \beta(a) \frac{\partial}{\partial a} + m\gamma (a) \frac{\partial}{\partial m}\right)\Gamma
\end{equation}
where
\begin{equation}
\beta(a) = \mu \frac{\partial a}{\partial \mu} = -ba^2(1 + ca + c_2a^2 + \ldots )
\end{equation}
and
\begin{equation}
m\gamma(a) = \mu\frac{\partial m}{\partial \mu} = mfa(1 + g_1 a+g_2a^2 + \ldots).
\end{equation}

The sums in eq. (1) can be organized using the functions
\begin{equation}
S_n(\xi) = \sum_{k=0}^\infty T_{n+k,k} \xi^k .
\end{equation}
Using eqs. (1,5) we can write
\begin{equation}
\Gamma = m^5 \sum_{n=0}^\infty a^n S_n\left( a\ln \frac{\mu}{m}\right).
\end{equation}
Substitution of eqs. (3,4,6) into eq. (2) leads to a set of nested equations with the boundary condition $S_n(0) = T_{n0}$; once $S_0, S_1 \ldots S_{n-1}$ are known, it is possible to solve for $S_n$.  The equation for $S_0$ is
\begin{equation}
S_0^\prime (\xi) - b\xi S_0^\prime (\xi) + 5 f S_0 (\xi) = 0
\end{equation}
so that
\begin{equation}
S_0(\xi) = T_{00} (1-b\xi)^{5f/b} .
\end{equation}
This is the LL contribution to $\Gamma$.

We now will write
\begin{equation}
\ln \left(\frac{\mu}{m} \right) = \ln \left(\frac{M}{m} \right) + \ln \left(\frac{\mu}{M} \right)
\end{equation}
where $M$ is chosen to be the pole mass of the $b$ quark.  We then have
\begin{equation}
\ln^k\left(\frac{\mu}{m} \right) = \sum_{r=0}^k\left( \begin{array}{c} k \\ r \end{array} \right) \Lambda^{k-r} L^r
\end{equation}
where
\begin{equation}\tag{11a,b}
\Lambda \equiv \ln \left(\frac{M}{m} \right), \quad L \equiv \ln \left(\frac{\mu}{M} \right). 
\end{equation}
Using eq. (10), eq. (1) can be rewritten as
\begin{equation}\tag{12}
\Gamma (a,m) = \sum_{n=0}^\infty A_n(a,m)L^n.
\end{equation}
Substitution of eq. (12) into eq. (2) leads to 
\begin{equation}\tag{13}
\sum_{n=0}^\infty \left[ (nA_n) L^{n-1} + \left( \beta(a) \frac{\partial}{\partial a} + m \gamma (a) \frac{\partial}{\partial m}\right) A_n L^n\right] = 0
\end{equation}
so that
\begin{equation}\tag{14}
A_n(a,m) = \frac{-1}{n}\left[ \beta (a) \frac{\partial}{\partial a} + m \gamma(a) \frac{\partial}{\partial m}\right] A_{n-1}(a,m).
\end{equation}
We now can define functions $\alpha (t)$, $\kappa (t)$ using
\begin{equation}\tag{15}
\frac{d\alpha (t)}{dt} = \beta(\alpha(t))\quad (\alpha (0) = a )
\end{equation}
\begin{equation}\tag{16}
\frac{d\kappa (t)}{dt} = \kappa(t)\gamma(\alpha(t)) \quad (\kappa (0) = m )
\end{equation}
so that by eqs. (14-16)
\begin{equation}\tag{17}
A_n(\alpha (t), \kappa(t)) = \frac{-1}{n}\frac{d}{dt} A_{n-1}(\alpha (t), \kappa (t))
\end{equation}
\begin{equation}\tag{18}
\hspace{3cm}
= \frac{(-1)^n}{n!}\frac{d^n}{dt^n}A_0(\alpha (t), \kappa(t)).
\end{equation}
Together, eqs. (12,18) lead to
\begin{equation}\tag{19}
\Gamma (\alpha (t), \kappa (t)) = \sum_{n=0}^\infty \frac{(-L)^n}{n!}\frac{d^n}{dt^n} A_0 (\alpha (t), \kappa (t))
\end{equation}
\begin{equation}\tag{20}
\hspace{5.6cm} = A_0 \left( \alpha \left( t + \ln \left(\frac{M}{\mu}\right)\right)\right), 
\kappa \left( t + \ln \left(\frac{M}{\mu}\right)\right),
\end{equation}
which upon setting $t = 0$ becomes
\begin{equation}\tag{21}
\Gamma(a,m) = A_0\left( \alpha \left( \ln \frac{M}{\mu}\right),  
\kappa \left( \ln \frac{M}{\mu}\right)\right).
\end{equation}

By eq. (15) we see that
\begin{equation}\tag{22}
\ln\left( \frac{M}{\mu}\right) = \int_a^{\alpha(\ln\frac{M}{\mu})} \frac{dx}{\beta(x)}
\end{equation}
and so
\begin{equation}\tag{23}
-1 = \frac{1}{\beta(\alpha(\ln \frac{M}{\mu}))} \left( \mu \frac{d\alpha(\ln\frac{M}{\mu})}{d\mu} \right) - \left( \mu \frac{1}{\beta(a)}\frac{da}{d\mu}\right)
\end{equation}
and thus, by eqs. (3,23)
\begin{equation}\tag{24}
\mu \frac{d\alpha}{d\mu}\left(\ln \frac{M}{\mu}\right) = 0.
\end{equation}

Together eqs. (15,16) show that
\begin{equation}\tag{25}
\frac{d\kappa}{d\alpha} = \frac{\kappa \gamma (\alpha)}{\beta(\alpha)} 
\end{equation}
and so
\begin{equation}\tag{26}
\int_m^{\kappa(\ln \frac{M}{\mu})} \frac{dx}{x} = \int_a^{\alpha(\ln \frac{M}{\mu})} dx \frac{\gamma(x)}{\beta(x)}.
\end{equation}
From eq. (26) we see that
\begin{align}\tag{27}
\frac{1}{\kappa(\ln \frac{M}{\mu})} & \mu \frac{d}{d\mu}\kappa \left( \ln \frac{M}{\mu}\right) - \frac{1}{m}\mu \frac{dm}{d\mu}\\
& = \frac{\gamma(\alpha (\ln \frac{M}{\mu}))}{\beta(\alpha (\ln \frac{M}{\mu}))} \mu 
\frac{d\alpha (\ln \frac{M}{\mu})}{d\mu}\nonumber \\
& \qquad - \frac{\gamma(a)}{\beta(a)}\mu \frac{da}{d\mu}.\nonumber 
\end{align}
Together, eqs. (3,4,24,27) show that
\begin{equation}\tag{28}
\mu\frac{d}{d\mu} \kappa \left( \ln \frac{M}{\mu}\right) = 0.
\end{equation}
As a result of eqs. (24,28), we see that the RG summed expression for $\Gamma(a,m)$ in eq. (21) is independent of $\mu$.  This means that when $M$, the pole mass, is expressed in units of some mass scale, and $\alpha(t)$ and $\kappa(t)$ have their values prescribed at that mass scale (necessarily from experiment) no mass scale ambiguity exists.

There are, however, RS ambiguities that can be parameterized by the coefficients $c_i (i \geq 2)$ in eq. (3) [4] and the coefficients $g_i(i \geq 1)$ in eq. (4) [5].   Following ref. [4], we can set
\begin{equation}\tag{29}
\frac{da}{dc_i} = B_i (a) = a^{i+1} \left( W_0^i +W^i_1 a+ W^i_2 a^2 + \ldots \right)
\end{equation}
and by using the equation
\begin{equation}\tag{30}
\left( \mu \frac{\partial}{\partial\mu} \frac{\partial}{\partial c_i} - \frac{\partial}{\partial c_i} \mu \frac{\partial}{\partial \mu}\right) a = 0
\end{equation}
we find that
\begin{align}\tag{31}
B_i (a) & = -b \beta(a) \int_0^a dx \frac{x^{i+2}}{\beta^2(x)}\\
& \approx a^{i+1} \left[ \frac{1}{i-1} - c \left(\frac{i-2}{i(i-1)}\right)a + \frac{1}{i+1}\left( c^2 \frac{i-2}{i} - c_2 \frac{i-3}{i-1}\right)a^2 + \ldots \right].\nonumber
\end{align}
We also take
\begin{equation}
\frac{\partial a}{\partial g_i} = 0, \quad \frac{\partial m}{\partial c_i} = m\Gamma^{ci}(a), \quad  \frac{\partial m}{\partial g_i} = m\Gamma^{gi}(a);\nonumber
\end{equation}
by using equations analogous to eq. (30) we find that
\begin{align}\tag{32}
\Gamma^{ci}(a) & = \frac{b\gamma(a)}{\beta(a)} B_i(a) + b \int_0^a dx \frac{x^{i+2}\gamma(x)}{\beta^2(x)}\\
& \approx \frac{f}{b} a^i \bigg[ \frac{-1}{i(i-1)} + 2 \left( \frac{c}{i(i+1)} - \frac{g_1}{(i+1)(i-1)}\right) a \nonumber \\
& \quad + \frac{1}{(i+2)} \left( \frac{2c_2}{i+1} - \frac{3c^2}{i+1} + \frac{4g_1c}{2} - \frac{3g_2}{i-1}\right) a^2 + \ldots \bigg]\nonumber 
\end{align}
and
\begin{align}\tag{33}
\Gamma^{gi} (a)& = f\int_0^a dx \frac{x^{i+1}}{\beta(x)} \\
& \approx \frac{f}{b} a^i \left[ - \frac{1}{i} + \left( \frac{c}{i+1}\right) a + \left(\frac{c_2-c^2}{i+2}\right) a^2 + \ldots \right].\nonumber
\end{align}

We now can examine the RS dependence of the RG summed expression for $\Gamma$ given in eq. (21).  We begin by noting that by eqs. (1,10,12), eq. (21) becomes
\begin{equation}\tag{34}
\Gamma(a,m) = \kappa^5 \left( \ln \frac{M}{\mu}\right) \sum_{n=0}^\infty \sum_{k=0}^n T_{nk} \alpha^n \left( \ln \frac{M}{\mu}\right)\ln^k \left(  \frac{M}{\kappa\left( \ln \frac{M}{\mu}\right)}\right).
\end{equation}

We now can use the equations
\begin{align}\tag{35a}
\frac{d\Gamma}{dc_i} = 0 &= \left( \frac{\partial}{\partial c_i}  + B^i (\alpha) \frac{\partial}{\partial \alpha} + \kappa \Gamma^{ci} (\alpha) 
\frac{\partial}{\partial\kappa}\right)\\
& \qquad\left( \kappa^5 \sum_{n=0}^\infty \sum_{k=0}^n T_{nk} \alpha^n \ln^k \left( 
\frac{M}{\kappa}\right)\right)\nonumber
\end{align}
\begin{equation}\tag{35b}
\frac{d\Gamma}{dg_i} = 0 = \left( \frac{\partial}{\partial g_i}  + \kappa \Gamma^{gi} (\alpha) \frac{\partial}{\partial \kappa}\right)
\left( \kappa^5 \sum_{n=0}^\infty \sum_{k=0}^n T_{nk} \alpha^n \ln^k \left( 
\frac{M}{\kappa}\right)\right)
\end{equation}
to arrive at
\begin{equation}\tag{36a-f}
\frac{\partial T_{00}}{\partial c_i} = \frac{\partial T_{10}}{\partial c_i} = \frac{\partial T_{11}}{\partial c_i} = \frac{\partial T_{21}}{\partial c_i}= \frac{\partial T_{22}}{\partial c_i} = 0, \quad 
\frac{\partial T_{20}}{\partial c_i} + \left(\frac{-5}{2} \frac{f}{6}\right)T_{00}\delta_{i2} = 0
\end{equation}
and
\begin{align}\tag{37a-f}
\frac{\partial T_{00}}{\partial g_i}& =  \frac{\partial T_{11}}{\partial g_i} = \frac{\partial T_{22}}{\partial g_i} = 0, \qquad \frac{\partial T_{10}}{\partial g_i} -5  \frac{f}{b}\delta_{i1} = 0,\nonumber \\
\frac{\partial T_{20}}{\partial g_i}& + 5 \frac{f}{b}\left[\left(\frac{T_{11}}{5} - T_{10} + cT_{00}\right) \delta_{i1} + \left(-\frac{T_{00}}{2}\right) \delta_{i2}\right] = 0\nonumber \\
\frac{\partial T_{21}}{\partial g_i}& -5 \frac{f}{b}T_{11}\delta_{i1} = 0 \nonumber
\end{align}
etc.  As a result of eqs. (36,37) we find that
\begin{align}\tag{38a-f}
 T_{00}&=\tau_{00}, \quad T_{10} =   \frac{5f}{b} \tau_{00} g_1 + \tau_{10}, \quad T_{11} = \tau_{11} \\
 T_{22}&=\tau_{22}, \quad T_{21} =    \frac{5f}{b} \tau_{11} g_1 + \tau_{21}, \nonumber \\
 T_{20}&= \left( \frac{5f}{2b}\right) \tau_{00} c_2 + \left[ \frac{1}{2} \left(\frac{5fg_1}{b}\right)^2 \tau_{00}\right]\nonumber \\
& \quad
 + \left[ \frac{5f}{2b} \tau_{00}\right]g_2 + \left[  \frac{5f}{b} \tau_{10} -  \frac{5fc}{2b} \tau_{00} -  \frac{f}{b} \tau_{11}\right]g_1 + \tau_{20}\nonumber
\end{align}
etc. In eq. (38), the $\tau_{nk}$ are constants of integration and are renormalization scheme invariants.  These constants are not all independent, as upon substitution of eqs. (1,3,4) into eq. (2) we find that 
\begin{align}\tag{39}
m^5 \sum_{n=0}^\infty \sum_{k=0}^n & T_{nk} \left\lbrace a^n k \ln^{k-1} \left(\frac{\mu}{m}\right) - ba^2 (1 + ca + c_2a^2 + \ldots) na^{n-1} \ln^k \left(\frac{\mu}{m}\right)\right. \\
&\left. + fa(1 + g_1 a + \ldots) \left( 5a^n \ln^k \left(\frac{\mu}{m}\right) - ka^n\ln^{k-1} \left(\frac{\mu}{m}\right)\right) \right\rbrace = 0\nonumber
\end{align}
from which follows, for example
\begin{equation}\tag{40}
n T_{nn} + \left( 5f - (n-1)b\right) T_{n-1,n-1} = 0
\end{equation}
(from terms in eq. (39) of order $a^n\ln^{n-1}\left(\frac{\mu}{m}\right)$), and 
\begin{equation}\tag{41}
n T_{n+1,n} + (5f - bn) T_{n,n-1} -nf T_{n,n} + \left(5fg_1 - bc(n-1)\right)T_{n-1,n-1} = 0
\end{equation}
(from terms in eq. (39) of order $a^{n+1}\ln^{n-1}\left(\frac{\mu}{m}\right)$).

By eqs. (38,40) we see that
\begin{equation}\tag{42a}
\tau_{11} + 5f \tau_{00} = 0
\end{equation}
\begin{equation}\tag{42b}
2\tau_{22} + (5f + b) \tau_{11} = 0;
\end{equation}
by eqs. (38,41,42a) it follows that
\begin{equation}\tag{42c}
\tau_{21} - f \tau_{11} + (5f-b)\tau_{10} = 0.
\end{equation}
Further relations between the RS constants $\tau_{nk}$ can be found in similar fashion.  Dependency on the RS dependent parameters $g_i(i \geq 1)$, $c_i (i \geq 2)$ cancels out in these relations.

There are two choices of RS that merit special attention.  In one scheme, the values of $c_i$, $g_i$ are chosen so that
\begin{equation}\tag{43}
T_{nk} = 0 \qquad (k < n),\qquad T_{nn} = \tau_{nn}.
\end{equation}
In this case, we have for example from eq. (38b)
\begin{equation}\tag{44}
g_1 = - \frac{b\tau_{10}}{5f\tau_{00}}
\end{equation}
so that $T_{10} = 0$.  Eq. (34) then reduces to 
\begin{equation}\tag{45}
\Gamma (a,m) = \kappa_1^5 \sum_{n=0}^\infty \tau_{nn}\alpha_1^n \ln^n \left(\frac{M}{\kappa_1}\right)
\end{equation}
where $\alpha_1$ and $\kappa_1$ are the functions introduced in eq. (15,16) using this RS evaluated at $\ln\left(\frac{M}{\mu}\right)$.  But by eq. (5), we see that eq. (45) is just a LL sum, so by eq. (8), we see that
\begin{equation}\tag{46}
\Gamma (a,m) = \kappa_1^5 \tau_{00} \left( 1 - b \alpha_1 \ln \frac{M}{\kappa_1}\right)^{5f/b}
\end{equation}
which is a very compact expression for $\Gamma$.

A second scheme in which one makes the choice
\begin{equation}\tag{47}
c_i = 0 ( i \geq 2), \qquad g_i = 0 (i \geq 1)
\end{equation}
leads to
\begin{equation}\tag{48}
\Gamma = \kappa_2^c \sum_{n=0}^\infty \sum_{k=0}^n \tau_{nk} \alpha_2^n \ln^k \left( \frac{M}{\kappa_2}\right)
\end{equation}
where now in eq. (48), by eqs. (22,26), $\alpha_2$ and $\kappa_2$ are simply given by
\begin{equation}\tag{49}
\ln \frac{M}{\mu} = \int_a^{\alpha_2} \frac{dx}{-bx^2(1+cx)}
\end{equation}
and
\begin{equation}\tag{50}
\int_m^{\kappa_2} \frac{dx}{x} = \int_a^{\alpha_2} dx \left( \frac{f}{-bx(1+cx)}\right).
\end{equation}

In eq. (9), $M$ was chosen to be the pole mass of the $b$ quark, but in fact it is an arbitrary mass parameter that is independent of the RS.  By eqs. (49,50)
\begin{equation}\tag{51a}
\qquad M \frac{d\alpha_2}{dM} = -b \alpha_2^2 (1 + c\alpha_2)
\end{equation}
\begin{equation}\tag{51b}
M \frac{d\kappa_2}{dM} = \kappa_2 f\alpha_2 \;,
\end{equation}
and so we have
\begin{equation}\tag{52}
M \frac{d\Gamma}{dM} = 0 = \left[ M \frac{\partial}{\partial M} + (-b\alpha_2^2) (1 + c\alpha_2) \frac{\partial}{\partial \alpha_2} + \kappa_2 f\alpha_2 \frac{\partial}{\partial \kappa_2}\right]\Gamma\;.
\end{equation}
Upon organizing the sums in eq. (48) as in eq. (6), eq. (52) becomes
\begin{align}\tag{53}
\kappa_2^5 \sum_{n=0}^\infty \bigg\{ S_n^\prime & (\xi) \alpha_2^{n+1} - b\alpha_2^2 (1 + c\alpha_2)\left( n S_n (\xi) \alpha_2^{n-1} + S_n^\prime (\xi) \xi \alpha_2^{n-1}\right) \nonumber \\
  &  + f\alpha_2 \left( 5 S_n (\xi) \alpha_2^n - S_n^\prime (\xi) \alpha_2^{n+1} \right)\bigg\} = 0 \nonumber
\end{align}
where $\xi = \alpha_2 \ln \left(\frac{M}{\kappa_2}\right)$.  Eq. (53) results in the nested equations for $n = 0,1,2 \ldots$
\begin{equation}\tag{54}
(1-b\xi) S_n^\prime + (5f-bn)S_n - (f + bc \xi)S_{n-1}^\prime - bc(n-1)S_{n-1} = 0
\end{equation}
with $S_n(0) = \tau_{n0}$ as the boundary condition.  The N$^n$LL contribution to the sum in eq. (48) is given by $S_n(\xi)$.  $\Gamma$ is determined by the RS invariants $b$, $c$, $f$ and $\tau_{n0}$.

The resummation that leads from eq. (1) to eq. (21) can also be applied to eq. (48).

\section{Discussion}

The arguments of ref. [8] have been extended to the amplitude $\Gamma$ for the decay $b \rightarrow u \ell^-\overline{\nu}_\ell$, which necessitates consideration of RS ambiguities that arise when the mass of the $b$ quark is renormalized.  It has been shown how when using a mass-independent RS, $\Gamma$ can be expressed in terms of the pole mass of the $b$ quark and a set of RS invariant parameters.  Two schemes are of particular interest; in one scheme $\Gamma$ is given by a LL expression (eq. (46)) while in a second scheme $\Gamma$ involves only the two-loop running coupling and the one-loop running mass (eq. (48)).

We hope to use the results presented above to perform a quantitative analysis of $\Gamma$.  We also plan to use this approach to examine RS ambiguities arising when there are multiple couplings, in thermal field theory [9], a constant external gauge field [10], and examine the decoupling of heavy quarks.

\section*{Acknowledgements}
Roger Macleod had a useful suggestion.

\end{document}